\documentclass[twocolumn,showpacs,preprintnumbers,amsmath,amssymb,a4paper,10pt]{revtex4}

\usepackage{graphicx}
\usepackage{bbm}
\usepackage{bm}
\graphicspath{{fig/}}

\pacs{71.27.+a,75.30.Ds,03.67.Lx}
 
\begin{document}

\title{The Quantum Compass Model on the Square Lattice} \author{Julien
\surname{Dorier},$^1$ Federico \surname{Becca}$^2$ and
Fr\'{e}d\'{e}ric \surname{Mila}$^1$} \affiliation{$^1$ Institut de
Th\'{e}orie des Ph\'{e}nom\`{e}nes Physiques, Ecole Polytechnique
F\'{e}d\'{e}rale de Lausanne (EPFL), CH-1015 Lausanne, Switzerland\\
$^2$ INFM-Democritos, National Simulation Centre, and SISSA I-34014
Trieste, Italy}

\date{\today}

\begin{abstract}
Using exact diagonalizations, Green's function Monte Carlo simulations
and high-order perturbation theory, we study the low-energy properties
of the two-dimensional spin-1/2 compass model on the square lattice
defined by the Hamiltonian $H = - \sum_{\bm{r}} \left( J_x
\sigma_{\bm{r}}^x \sigma_{\bm{r} + \bm{e}_x}^x + J_z \sigma_{\bm{r}}^z
\sigma_{\bm{r} + \bm{e}_z}^z \right)$. When $J_x\ne J_z$, we show
that, on clusters of dimension $L\times L$, the low-energy spectrum
consists of $2^L$ states which collapse onto each other
exponentially fast with $L$, a conclusion that remains true
arbitrarily close to $J_x=J_z$. At that point, we show that an even
larger number of states collapse exponentially fast with $L$ onto the
ground state, and we present numerical evidence that this number is
precisely $2\times 2^L$. We also extend the symmetry analysis of the
model to arbitrary spins and show that the two-fold degeneracy of all
eigenstates remains true for arbitrary half-integer spins but does not
apply to integer spins, in which cases eigenstates are generically non
degenerate, a result confirmed by exact diagonalizations in the spin-1
case. Implications for Mott insulators and Josephson junction arrays
are briefly discussed.
\end{abstract}

\maketitle

\section{Introduction}

Building on the deep understanding of the Heisenberg and other models
of magnetism, it is a very common practice to describe discrete
degrees of freedom as pseudo-spins, with the hope to gain insight from
the form of the resulting magnetic model. A well-known example of
considerable current interest shows up in the context of Mott
insulators with orbital degeneracy: In an octahedral environment, the
degeneracy of the $d$ electrons is only partially lifted, and the
remaining orbital degree of freedom is often described as a spin-1/2
or 1 for $e_g$ and $t_{2g}$ electrons, respectively~\cite{kugel1,kugel2}. 
However, the situation is in general
less simple that one might hope. Indeed, as already emphasized by
Kugel and Khomskii, the symmetry of pseudo-spin Hamiltonians is in
general much lower than SU(2), and there are cases where the
properties of the resulting model are poorly understood. This is in
particular the case of models where the anisotropy of the coupling in
spin space is related to the orientation of the bond in real
space. The simplest version of such a model on the square lattice is
defined by the Hamiltonian
\begin{equation}
  H = - J \sum_{\bm{r}} \left(\ \tau_{\bm{r}}^x \tau_{\bm{r} +
  \bm{e}_x}^x + \ \tau_{\bm{r}}^z \tau_{\bm{r} + \bm{e}_z}^z \right),
  \label{intro1}
\end{equation}
where $\tau_{\bm{r}}^\alpha$ are the $x$ and $z$ components of a pseudo-spin
operator.
By analogy with the dipolar coupling between compass needles, this
model has been called the compass model by Kugel and
Khomskii~\cite{kugel2}. Realistic models of orbital degeneracy are usually more
complicated in several respects. In particular, the spins and
pseudo-spins are usually coupled. Pure orbital models can be of direct
relevance though if the spins order ferromagnetically, as recently
argued by Mostovoy and Khomskii in the context of NaNiO$_2$~\cite{mostovoy}. The
precise symmetry is also usually more complicated than the simple case
of this Hamiltonian, but we will nevertheless concentrate on that
model, considering it as a minimal model rather than a realistic one.

Interestingly, such models have appeared in other contexts as
well. First, even in Mott insulators without orbital degeneracy, extra
degrees of freedom can appear if the system is frustrated and, for instance,
consists of spin-1/2 coupled triangles, like in the trimerized kagome
lattice, in which case the chirality that keeps track of the extra
degeneracy of each triangle plays a role similar to that of
orbitals~\cite{mila}. In a magnetic field, this model has been
predicted to exhibit a magnetization plateau at 1/3~\cite{cabra1}, and
the low-energy properties inside the plateau can be described by a
kind of compass model~\cite{cabra2}.

More recently, the model of Eq.~(\ref{intro1}) with anisotropic
couplings along the $x$ and $z$ directions has been proposed by 
Dou\c{c}ot and collaborators in the context of Josephson junction
arrays~\cite{doucot}.

Despite its deceptive simplicity, the model of Eq.~(\ref{intro1}) is a
formidable challenge, in many respect comparable to very frustrated
magnets. To see this, let us follow Ref.~\cite{nussinov1} and
consider the classical version of the model, in which spins are
considered as classical vectors. In that case, as shown by Nussinov 
{\it et al.}, the ground state is highly degenerate, as in 
very frustrated magnets. First of all, all ferromagnetic states are degenerate,
regardless of the relative orientation of the spins with respect to
the lattice, as can be easily checked from Eq.~(\ref{intro1}).  In
addition, from any ferromagnetic state, one can construct other states
by flipping all spins of a $z$ column with respect to an $x$ mirror,
or equivalently by flipping all spins of an $x$ line with respect to a
$z$ mirror. Since all these operations can be performed simultaneously
and in any order, they generate a discrete degeneracy of order $2^L$.

The effects of thermal fluctuations on the classical model have been
convincingly identified by analytical and numerical
approaches. Nussinov {\it et al.} have shown that an order by disorder
mechanism is expected to lift the rotational degeneracy and to select
states in which the spins point along the $x$ or $z$ axis, leading to
a nematic ground state since lines or columns of spins are still free
to flip. Using extensive Monte Carlo simulations, Mishra {\it et al.} have
shown that the two possible orientations along $x$ or $z$ lead to an
effective Ising order parameter, and that the model undergoes a finite
temperature phase transition of the Ising type~\cite{mishra}.

On the other hand, the understanding of the spin-1/2 version of the model
is still preliminary. Most of the results have been obtained 
by Dou\c{c}ot {\it et al.}~\cite{doucot} in their analysis of a generalized 
version of the model of Eq.~(\ref{intro1}) defined by the Hamiltonian
\begin{equation}
  H = - \sum_{\bm{r}} \left( J_x \ \tau_{\bm{r}}^x \tau_{\bm{r} +
  \bm{e}_x}^x + J_z \ \tau_{\bm{r}}^z \tau_{\bm{r} + \bm{e}_z}^z
  \right),
  \label{intro2}
\end{equation}
in which the couplings along the $x$ and $z$ directions can take
different values.  Using elegant symmetry arguments, Dou\c{c}ot {\it et al.}
have shown that all eigenstates must be at least two-fold degenerate. They have
also shown that in the strongly asymmetric case ($J_x/J_z \ll 1$ or 
$J_z/J_x \ll 1$), the $2^L$ states that evolve adiabatically from the $2^L$
ground states of the decoupled Ising chain case ($J_x=0$ or $J_z=0$) should
collapse onto each other.  This perturbative argument does not
apply close to $J_x=J_z$ though, and whether this remains true in the
isotropic case could not be decided. Note that this is an important
issue in the context of quantum bits in which they came accross this
model since the presence of a gap would help to protect the
q-bits. More recently, Nussinov and Fradkin~\cite{nussinov2} have shown that 
these models are dual to models of $p+ip$ superconducting arrays, and have
discussed general properties of order parameters and phase
transitions.

In this paper, we concentrate on the zero-temperature properties of
the quantum version of the model. We start by a semi-classical
analysis of the model of Eq.~(\ref{intro1}) and show that, as in many
frustrated magnets, quantum fluctuations essentially have the same
effect as thermal fluctuations regarding the lifting of classical
degeneracy (section~\ref{sec_spinwave}). We then turn to an extensive
analysis of the model of Eq.~(\ref{intro2}) in the spin-1/2 case. As we
shall see, there is no minus sign problem in the Green's function
implementation of quantum Monte Carlo, which allows one to study the
ground state properties of very large clusters, i.e., up to 17 $\times$ 17. 
Combined with the large number of symmetries, hence of different
symmetry sectors that can be studied independently, this allowed us to
reach definite conclusions regarding the low energy spectrum,
conclusions that agree with high order perturbation theory. These
results are presented in section~\ref{sec_spindemi}. Finally, the
symmetry analysis of Dou\c{c}ot {\it et al.} is extended to larger spins in
section~\ref{sec_spin1}, with the conclusion that integer and
half-integer spins behave once more quite differently. Some
implications of the present results are discussed in the last section
of the paper.

\section{\label{sec_spinwave}Semi-classical compass model}

In order to have a first insight into the properties of the quantum
version of the compass model, we have performed a spin-wave analysis
in the symmetric case defined by Eq.~(\ref{intro1}). In that respect,
it is useful to emphasize that, as noticed in
Refs.~\cite{nussinov1,mishra}, the degeneracy is partly accidental and
partly due to symmetry.  Indeed, in addition to the lattice
translational symmetries, this model has two types of discrete
symmetries: (i) The $Q_i$ transformation which flips the $z$ component
of all the spins of the column $r_x = i$, and the $P_j$
transformations which flip the $x$ component of all spins of the line
$r_z = j$. Note that the transformations $Q_i$ (resp. $P_j$) could be
seen as rotations of all the spins of the column (resp. line) about
the $\bm{e}_x$ axis (resp. $\bm{e}_z$) by an angle $\pi$. (ii) The
simultaneous rotation $R_y ( \frac{\pi}{2} )$ of all spins {\it and}
of the lattice about an $\bm{e}_y$ axis by an angle
$\frac{\pi}{2}$. Starting from any state, these symmetries generate
new states with exactly the same classical energy and the degeneracy
associated to these symmetries cannot be lifted by thermal
fluctuations~\cite{mishra}. However, the ground state also has an
accidental degeneracy: All ferromagnetic states are degenerate
regardless of the angle between the spins and the lattice. This
degeneracy is not related to a symmetry since the model is not
rotationally invariant.  Accordingly, it has been found that the
thermal fluctuations partially lift this degeneracy via an
order-by-disorder mechanism, favouring the $2\times2^L$ ground states
with all spins parallel to $\bm{e}_x$ or $\bm{e}_z$ and implying a
directional ordering of the spins. These favoured states are the
uniform state with all spins in the direction $\bm{e}_x$ and all the
states obtained by applying the symmetries $R_y ( \frac{\pi}{2} )$,
$P_j$ and $Q_i$. So, as anticipated, only the accidental degeneracy is
lifted by thermal fluctuations.

\begin{figure}[th]
        \includegraphics[width=0.48\textwidth]{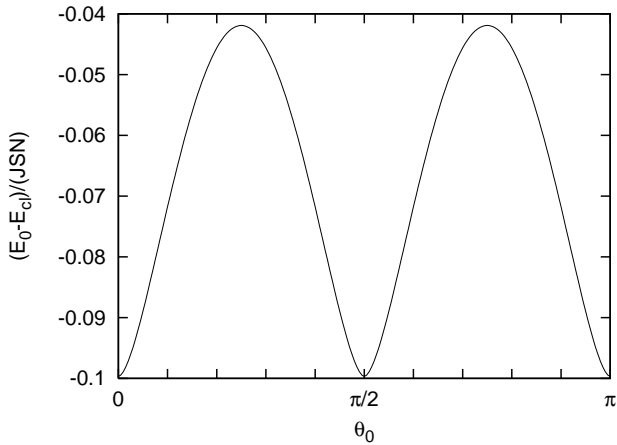}
        \caption{\label{fig_spinwave_Efond}Ground-state energy of the
        ferromagnetic states defined by $\bm{S_r} = \bm{S} ( \theta_0
        ) = S \cos \theta_0 \bm{e}_x + S \sin \theta_0 \bm{e}_z$
        including zero-point energy as a function of the angle
        $\theta_0$. $E_{cl}=-JNS^2$ is the classical ground-state energy.}
\end{figure}

The same idea applies to quantum fluctuations. Starting from an
arbitrary ground state, one can bring it back into a ferromagnetic
ground state applying only symmetry operations. But applying symmetry
operations does not change the form of the Hamiltonian. The
fluctuations around both states will thus have exactly the same form.
Then it is sufficient to do the spin-wave expansion around the uniform
classical ground states $\bm{S_r} = S \cos \theta_0 \bm{e}_x + S \sin
\theta_0 \bm{e}_z$. To linear order in $1/S$, the energy can be
brought into the form
\begin{equation}
H=E_0+\frac{1}{2}\sum_q \omega_q(\theta_0),
\end{equation}
with $\omega_q(\theta_0)=4JS\sqrt{1-\cos^2 \theta_0 \cos k_x
  -\sin^2\theta_0 \cos k_z }$, and the resulting ground-state energy is
plotted in Fig.~\ref{fig_spinwave_Efond}.
As in the classical case with thermal fluctuations, the figure clearly
shows that the angles $\theta_0 = 0, \frac{\pi}{2}, \pi,
\frac{3\pi}{2}$ are selected by quantum fluctuations since they
minimize the energy.  Applying the symmetries $Q_i$, $P_j$ and $R_y (
\frac{\pi}{2} )$ to these states gives $2 \times 2^L$ equivalent
favoured states, corresponding to $2^L$ states parallel to $\bm{e}_x$ and
$2^L$ states parallel to $\bm{e}_z$. So, as in the classical case, there is
a directional ordering of the ground state.

As is many frustrated magnets like the $J_1-J_2$ model on the square
lattice for $J_2/J_1=1/2$, this calculation is not fully consistent
since the correction to the magnetization diverges~\cite{chandra}. 
In the present
case, the divergence comes from a line of zero energy along the $k_x=0$
direction when $\theta_0=0,\pi$ and along the $k_z=0$ direction when
$\theta_0=\frac{\pi}{2},\frac{3\pi}{2}$. We have pushed the expansion
to next order in 1/S and checked that a self-consistent mean-field
decoupling of the 4-boson terms suppresses the divergence. At this
approximation, the spectrum becomes gapped, which does not violate any
general theorem since the model does not possess rotational symmetry,
and long-range order is preserved even for spin-1/2.

In the anisotropic case ($J_x\neq J_z$), discrete rotational symmetry 
of the Hamiltonian as well as continuous rotational symmetry of the 
ground state are lost, leading to only the $2^L$ ground states with rows
of parallel spins along 
the $x$ axis for $J_x>J_z$ and columns of parallel spins along the $z$ 
axis for $J_x<J_z$. Since these $2^L$
classical ground states are related by symmetries of the Hamiltonian,
quantum fluctuations cannot lift this degeneracy. 

\section{\label{sec_spindemi}quantum compass model: spin-1/2}

In this section, we turn to the spin-1/2 case for which we write the
Hamiltonian
\begin{equation}
 H = - \sum_{\bm{r}} \left( J_x \sigma_{\bm{r}}^x \sigma_{\bm{r} +
   \bm{e}_x}^x + J_z \sigma_{\bm{r}}^z \sigma_{\bm{r} + \bm{e}_z}^z
   \right) \label{IntroQuantique1},
\end{equation}
where $\sigma_{\bm{r}}^x$ and $ \sigma_{\bm{r}}^z$ are Pauli matrices acting 
on the spin at site $\bm{r}$. In the following, we will use the 
parametrization of the
exchange integrals $J_x = J \cos \theta$ and $J_z = J \sin \theta$
with $\theta \in [ 0, \frac{\pi}{2} ]$, and we will study the model on
$N$-site square clusters of dimension $L\times L$.

Taking all symmetries into account, exact diagonalizations could be
performed up to $L=5$. As we shall see, this is not sufficient to draw
conclusions regarding the degeneracy of the ground state in the
thermodynamic limit close to $J_x=J_z$. However, this model has the
very interesting property that all non-diagonal matrix elements are
negative. This has allowed us to implement the Green's function Monte
Carlo algorithm~\cite{calandra}, which gives access to the
ground-state energy in a given symmetry sector, and
to reach clusters up to $L=17$. Besides, one can choose
quantum numbers so that all relevant low-energy states are ground
states of a given symmetry sector. To see how this works, let us look
more closely at the symmetries of the model.

In addition to the lattice translation symmetries, the
Hamiltonian~(\ref{IntroQuantique1}) has another type of discrete
symmetries.  The first one corresponds to the operators $Q_i = \prod_j
\sigma_{i, j}^x$, which are the products of the $\sigma_{\bm{r}}^x$ on
one column $( r_x = i )$. These operations correspond to a rotation by
an angle $\pi$ about the $\bm{e}_x$ axis of all the spins of a given
column: $Q_i^{- 1} \sigma_{i, j}^{y, z} Q_i = - \sigma_{i, j}^{y, z}$
and $Q_i^{- 1} \sigma_{i, j}^x Q_i = \sigma_{i, j}^x$. The second one
corresponds to the operators $P_j = \prod_i \sigma_{i, j}^z$, which
are the products of the $\sigma_{\bm{r}}^z$ on one line $( r_z = j )$,
and which correspond to a rotation by an angle $\pi$ about the
$\bm{e}_z$ axis of all the spins of a given line: $P_j^{- 1}
\sigma_{i, j}^{x, y} P_j = - \sigma_{i, j}^{x, y}$ and $P_j^{- 1}
\sigma_{i, j}^z P_j = \sigma_{i, j}^z$. In the isotropic limit
$J_x=J_z$, there is one more discrete symmetry: the global rotation
$R_y ( \frac{\pi}{2} )$ of all spins and lattice about the $\bm{e}_y$
axis by an angle $\frac{\pi}{2}$.

As emphasized by Dou\c{c}ot {\it et al.}, the $P_j$'s commute with each other,
as well as the $Q_i$'s, but $[ Q_i, P_j ] \neq 0$ $\forall \ i,
j$. This has two remarkable consequences: First of all, all
eigenstates must be two-fold degenerate. Besides, we can choose either
the $P_j$'s or the $Q_i$'s to define symmetry sectors in which the
Hamiltonian can be independently diagonalized. Since $P_j^2 = 1$ and
$P_j^{\dag} = P_j$, the eigenvalues of $P_j$ are $p_j = \pm 1$, the
same being true for the $Q_i$'s.  Thus the Hamiltonian can be
diagonalized in the symmetry sectors characterized by the set
$(p_1,\cdots,p_L)$, $p_i=\pm 1$, or alternatively in the sectors
defined by the eigenvalues of $Q_i$ and characterized by the set
$(q_1,\cdots,q_L)$, $q_i=\pm 1$.

To see how this works, let us start from the
trivial case $J_x = 0$. The model then consists of a set of decoupled
Ising columns with eigenstates $| \bm{m} \rangle = |m_{1, 1} \rangle
\otimes |m_{1, 2} \rangle \otimes \cdots \otimes |m_{L, L} \rangle$,
where $m_{i,j}=\pm 1$ is the eigenvalue of $\sigma_{i, j}^z$. The
ground state manifold contains $2^L$ states defined by $m_{i, 1} =
\cdots = m_{i, L} = \pm 1$, $i=1,...,L$. Clearly, all $|
\bm{m}\rangle$ states are eigenstates of the $P_j$'s. Now, in any of
the ground state, all $p_j$'s are equal since all lines are identical,
and the ground state manifold consists of $2^{L - 1}$ states in the
sector $p_1 = \cdots = p_L = + 1$ and $2^{L - 1}$ states in the sector
$p_1 = \cdots = p_L = - 1$.

The classification according to the $Q_i$'s is quite different.
First, note that the states $\{ |\bm{m} \rangle \}$ are not
eigenstates of $Q_i$. However, if we denote by $|\uparrow \rangle_i$
(resp. $|\downarrow \rangle_i$) the ground state of column $i$ with all
the spins up (resp. down), then one can define two new ground states
by $| + \rangle_i = \frac{1}{\sqrt{2}} \left( | \uparrow \rangle_i + |
\downarrow \rangle_i \right)$ and $| - \rangle_i = \frac{1}{\sqrt{2}}
\left( | \uparrow \rangle_i - | \downarrow \rangle_i \right)$. These new
states are eignestates of $Q_i$ with eigenvalues $\pm 1$
respectively. So $q_i$ can be either $-1$ or $+1$ for each column
independently, and the ground state manifold has one member in each of
the $2^L$ sectors $(q_1,\cdots,q_L)$, $q_i=\pm 1$.

The case $J_z = 0$ is connected to the case $J_x = 0$ by the rotation
symmetry $R_y ( \frac{\pi}{2} )$. Since $R_y ( \frac{\pi}{2} )^{-1}
P_j R_y ( \frac{\pi}{2} )=(-1)^L Q_{L-j}$ and $R_y ( \frac{\pi}{2}
)^{-1} Q_i R_y ( \frac{\pi}{2} )= P_i$, the symmetry sectors are
interchanged. So the ground state is made up of $2^L$ states in each
$2^L$ sectors $( p_1, \cdots, p_L ) = ( \pm 1, \cdots, \pm 1 )$.

So, as announced earlier, to determine the structure of the low-energy
spectrum when going away from $J_x=0£$ or $J_z=0$, it is always
possible to choose the quantum numbers so that each state is the
ground state of a given symmetry sector.

\subsection{Exact diagonalization}

\begin{figure}[th]
        \includegraphics[width=0.48\textwidth]{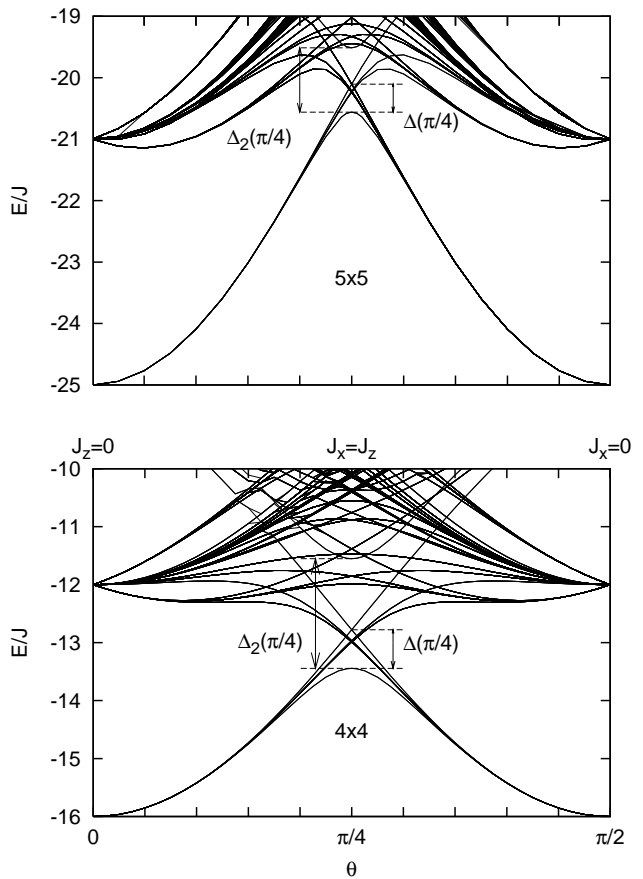}
        \caption{\label{fig_plot4x4_5x5_theta}Energy of the low-lying
states versus anisotropy parameter $\theta$ for $4\times 4$ and
$5\times 5$ lattices obtained by exact diagonalization ($J_x=J \cos
\theta, J_z=J \sin \theta$).}
\end{figure}

The spectrum of the Hamiltonian~(\ref{IntroQuantique1}) versus the
asymmetry parameter $\theta$ has been determined for $L=2,3,4$ and $5$
using exact diagonalization. Fig.~\ref{fig_plot4x4_5x5_theta}
presents the low energy levels for $L = 4$ and $L = 5$.
As expected, when $J_x,J_z\neq0$, the ground state is two-fold degenerate 
with one state in the
sector $p_1= \cdots = p_L = + 1$ and the other in the sector 
$p_1 = \cdots = p_L = -1$. For $0<J_z<J_x$, the $2^L$-fold degeneracy of the 
ground state at $\theta = 0$ (i.e., $J_z = 0$) is lifted by the $J_z$ term 
of the Hamiltonian. The lowest of these states is in the sector $p_1 =\cdots
= p_L = \pm 1$ and the highest in the sector $p_i = \pm (-1)^i$.  The gap
between these two states is denoted by $\Delta(\theta)$. As we will
see in the following, $\Delta(\theta)$ goes to zero for all $\theta$ in
the thermodynamic limit.

\begin{figure}[th]
        \includegraphics[width=0.48\textwidth]{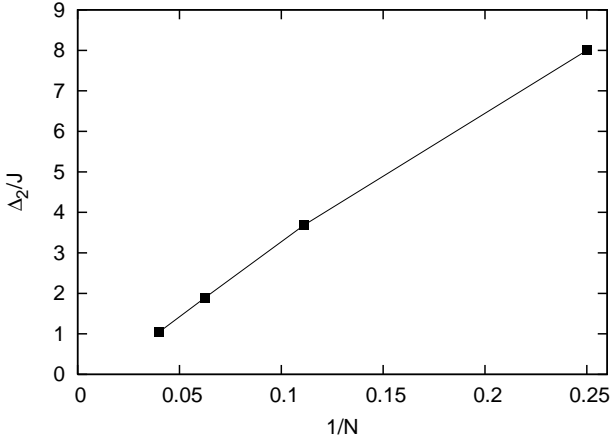}
        \caption{\label{fig_plotGap-N_1111}Gap $\Delta_2$ (see text)
as a function of the number of sites $N$.}
\end{figure}

In the symmetric case ($\theta = \frac{\pi}{4}$, i.e., $J_x = J_z$),
the adiabatic continuation of the degenerate ground states of the
$J_x=0$ and $J_z=0$ cases generates only $2 \times 2^L - 2$ states,
and not $2\times 2^L$ as one might naively expect from the
semi-classical case. The reason is that the lowest pair of state is
common to the two families of states coming from $J_x=0$ and $J_z=0$,
while all other states cross at $J_x=J_z$ (see
Fig.~\ref{fig_plot4x4_5x5_theta}).  This does not mean however that
the low energy sector has only $2\times 2^L-2$ states for very large
systems. In fact, another pair of states is decreasing quite fast
toward the ground state as a function of the size. Since these states
correspond to the first two-fold degenerate excited state of the
sectors $p_1 = \cdots = p_L = 1$ and $p_1 =\cdots = p_L = - 1$, one
can keep track of their energy for all sizes.  The gap between the
ground state and these states, denoted by $\Delta_2$, is plotted in
Fig.~\ref{fig_plotGap-N_1111} as a function of $1/N$. These results
are indeed consistent with a vanishing of this gap in the
thermodynamic limit~\cite{footnote}.

\subsection{Perturbation theory}

Before discussing the results obtained for large clusters with Green's
function quantum Monte Carlo, let us see what perturbation theory
predicts regarding the scaling of the gap $\Delta$ in the limit $J_z
\ll J_x$.  To this purpose, the Hamiltonian is written as
\begin{equation}
  H = H_0 + V,
\end{equation}
where $H_0 = - J_x \sum_{\bm{r}} \sigma_{\bm{r}}^x \sigma_{\bm{r} +
\bm{e}_x}^x$ and $V = - J_z \sum_{\bm{r}} \sigma_{\bm{r}}^z
\sigma_{\bm{r} + \bm{e}_z}^z$.

Since each term of $V$ flips a pair of spins on neighboring chains, it
is necessary to apply the perturbation at least $L$ times on an
$L\times L$ lattice to flip all spins of a pair of neighboring chains
and reach another ground state.  So the gap is expected to scale as
$\Delta/J_x=a (bJ_z/J_x)^L$.  One can be more precise though and
determine the constants $a$ and $b$ from high order perturbation
theory.  To first significant order in $J_z/J_x$, the gap $\Delta$ is
given by the gap between lower and higher eigenvalues of the effective
Hamiltonian
\begin{equation}
  H_{\text{eff}}^{( L )} = \sum_{\nu_1 > 0} \cdots \sum_{\nu_{L -
  1} > 0} \frac{P_0 V P_{\nu_1} V \cdots V P_{\nu_{L - 1}} V P_0}{(
  E_0 - E_{\nu_1} ) \cdots ( E_0 - E_{\nu_{L - 1}} )},
\end{equation}
where $P_{\nu}$ is the projector on the $\nu$-th eigenspace of $H_0$.
This eigenspace has an energy $E_{\nu} = - J_x N + 4 \nu J_x$, where
$\nu$ can take the values $\nu = 0, 1, \cdots, L/2$ when $L$ is even
and $\nu = 0, 1, \cdots, (L - 1)/2$ when $L$ is odd.
Let $\{ | \nu, k \rangle : k = 0, 1, \cdots \}$ be a basis of the
$\nu$-th eigenspace of $H_0$. We have to evaluate 
\begin{eqnarray*}
  &  &\hspace{-10mm} \langle 0, l|V P_{\nu_1} V \cdots V P_{\nu_{L - 1}} V| 0, k \rangle\\
  & = & ( - J_z )^L \sum_{i_1, j_1} \cdots \sum_{i_L, j_L} \langle 0, l|
  \sigma_{i_1, j_1}^z \sigma_{i_1, j_1 + 1}^z P_{\nu_1} \\
  & &\cdots P_{\nu_{L - 1}}
  \sigma_{i_L, j_L}^z \sigma_{i_L, j_L + 1}^z |0, k \rangle.
\end{eqnarray*}
For $L>2$, the only contribution to $\Delta$ arise when the $\sigma$
product flips two neighboring lines with $r_z = j$ and $r_z = j +
1$. It corresponds to $j_1 = \cdots = j_L = j$ and $i_1=p ( 1
),\cdots, i_L=p ( L )$ with $p \in S_L$ and $S_L$ the set of $L!$
permutations. The ground state $|0, k \rangle$ contains only
ferromagnetic lines in the $x$ or $-x$ direction. So, for $m<L$, the
state $\sigma_{p ( m + 1 ), j}^z \sigma_{p ( m + 1 ), j + 1}^z \cdots
\sigma_{p ( L ), j}^z \sigma_{p ( L ), j + 1}^z |0, k \rangle$ cannot
be in the ground state manifold of $H_0$, but it must be in one of the
excited eigenspaces, say the $\nu_m ( p )$-th. Then we can write:
\begin{eqnarray*}
  & &\hspace{-10mm} \langle 0, l|V P_{\nu_1} V \cdots V P_{\nu_{L -
  1}} V| 0, k \rangle\\
  & = & ( - J_z )^L \sum_{p \in S_n} \sum_j
  \langle 0, l| \sigma_{p ( 1 ), j_{}}^z \sigma_{p ( 1 ), j_{} + 1}^z\\ 
  & &\cdots \sigma_{p ( L ), j}^z \sigma_{p ( L ), j + 1}^z |0, k
  \rangle \prod_{m = 1}^{L - 1} \delta_{\nu_m, \nu_m ( p )}.
\end{eqnarray*}
Using $E_{\nu} = E_0 + 4 \nu J_x$ and rearranging the $\sigma^z$'s, we
get:
\begin{eqnarray*}
  & &\hspace{-10mm} \langle 0, l|H_{\text{eff}}^{( L )} |0, k
  \rangle\\ 
  & = & - \frac{J_z^L}{( 4 J_x )^{L - 1}} \sum_{p \in S_n}
  \frac{1}{\nu_1 ( p )_{} \cdots \nu_{L - 1} ( p )}\\ 
  & &\langle 0, l|
  \sum_j \sigma_{1, j_{}}^z \sigma_{1, j_{} + 1}^z \cdots \sigma_{L,
  j}^z \sigma_{L, j + 1}^z |0, k \rangle.
\end{eqnarray*}
Next, we note that, in terms of the pseudo-spin $\tau_j^z=\sigma_{1,
j}^z \cdots \sigma_{L, j}^z $, the operator $\sum_j \sigma_{1, j_{}}^z
\sigma_{1, j_{} + 1}^z \cdots \sigma_{L, j}^z \sigma_{L, j + 1}^z$ is
nothing but the Hamiltonian of the one-dimensional Ising model.  So,
the gap between the lowest and highest eigenvalues is $\lambda_{\max}
- \lambda_{\min} = 2 L$ when $L$ is even and $\lambda_{\max} -
\lambda_{\min} = 2 ( L - 1 )$ when $L$ is odd.  Finally, if we define
$P(L)$ by
\begin{equation}
P ( L ) = \sum_{p \in S_n} \frac{1}{\nu_1 ( p
)_{} \cdots \nu_{L - 1} ( p )},
\end{equation}
the gap $\Delta$ becomes
\begin{equation}
\Delta /J_x= \left\{ \begin{array}{ll}
     8LP(L)(J_z/4 J_x )^{L} & \text{if } L \text{ is
     even}\\\\
      8(L-1)P(L)(J_z/4 J_x )^{L} & \text{if } L \text{
     is odd}.
   \end{array} \right. 
\end{equation}
The dominant behaviour of $P ( L )$ has been determined 
numerically (see Fig.~\ref{fig_perturbation_Pn}).  It turns out that
$LP(L)\simeq \exp(0.754 L -0.694)$, which leads to:
\begin{figure}[th]
        \includegraphics[width=0.48\textwidth]{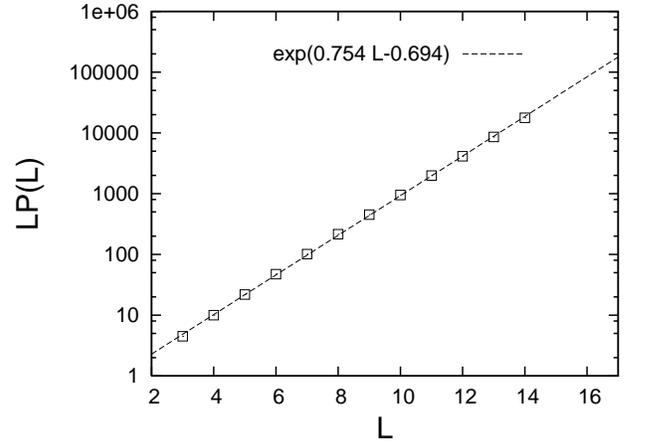}
        \caption{\label{fig_perturbation_Pn}log-linear plot of $LP(L)$
        versus the linear lattice size $L$. The squares are the
        results obtained numerically, and the dotted line is an
        exponential fit.}
\end{figure}
\begin{equation}
  \Delta/J_x = \left\{ \begin{array}{ll} 3.997  \left( 0.531
  J_z/J_x \right)^L & \text{ if } L \text{ even}\\\\
  3.997 (1-1/L)
  \left( 0.531 J_z/J_x \right)^L  & \text{ if }
  L \text{ odd.}
  \end{array} \right.
\end{equation}
It has the correct $\left( J_z/J_x \right)^L$ behavior which
corresponds to the exact result for the two-line system~\cite{doucot}. 
So, when $(J_z/J_x) \ll 1$, this approximation
predicts $\Delta \rightarrow 0$ in the thermodynamic limit. Moreover,
since $0.531 (J_z/J_x) < 1$ even for $J_z/J_x=1$, it seems
likely that this scaling will remain true up to the symmetric limit
$J_x=J_z$. As we shall see, this is confirmed by the Green's function
Monte Carlo results.

\subsection{Green's function Monte Carlo}

If the Hamiltonian of a model has only non-povitive off-diagonal
matrix elements, which is the case here, the Green's function Monte
Carlo method~\cite{calandra} allows one to calculate the ground-state
energy of a given symmetry sector by using a stochastic approach. The
algorithm we have used is the implementation with a fixed number of
walkers described in detail by Calandra and Sorella~\cite{calandra},
and the guiding function is given by:
\begin{equation}
  | \psi_G \rangle = \exp \left( \frac{1}{2} \sum_{\bm{r}, \bm{r}'}
  v_{\bm{r}, \bm{r}'} \sigma_{\bm{r}}^z \sigma_{\bm{r}'}^z \right)
  |F_x \rangle,
\end{equation}
where $|F_x \rangle$ is the ferromagnetic state in $x$ direction such
that $\sigma_{\bm{r}}^x |F_x \rangle = |F_x \rangle$
$\forall\bm{r}$. The parameters $v_{\bm{r}, \bm{r}'}$ were determined
in order to minimize the energy of the guiding function.  The ground
state energy has been determined in each sector $(p_1,\cdots,p_L)$
separately, which corresponds to the energy of the first $2^L$ states
of the full Hilbert space (the states coming from the limit $J_z=0$
ground state in Fig.~\ref{fig_plot4x4_5x5_theta}), and gives access to
the gap $\Delta(\theta)$.

\begin{figure}[th]
        \includegraphics[width=0.48\textwidth]{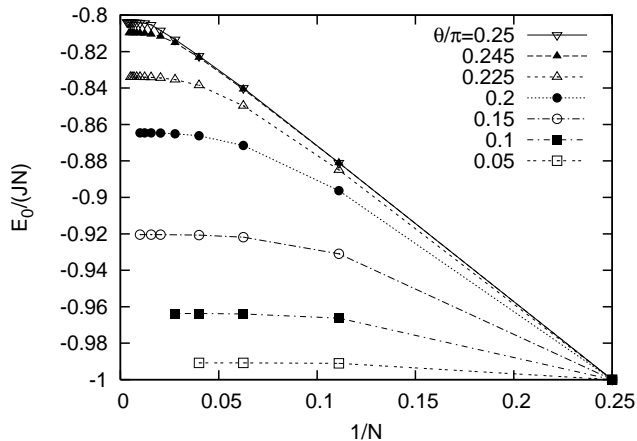}
        \caption{\label{fig_mc_plotEfond-N}Ground-state energy $E_0$
versus the number of sites $N$ for various values of the asymmetry parameter
$\theta$.}
\end{figure}

Let us first discuss the scaling of the energy per site as a function
of the system size. As can be seen in Fig.~\ref{fig_mc_plotEfond-N},
the energy per site is strongly size dependent up to a certain size
($8\times 8$ for $J_x=J_z$), and is very little size dependent for
larger clusters. This indicates that strong finite-size effects are to
be expected, especially close to the symmetric point, justifying the
use of quantum Monte Carlo to get information on large clusters.

\begin{figure}[th]
        \includegraphics[width=0.48\textwidth]{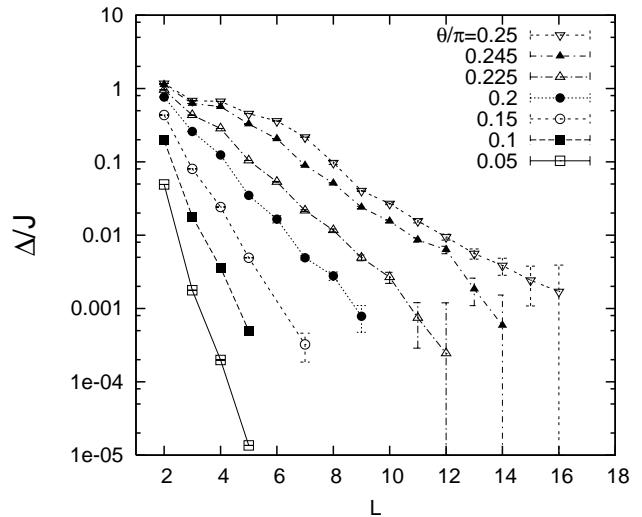}
        \caption{\label{fig_mc_plotGap-N_theta_excite}Gap $\Delta$
versus linear lattice size $L$ for various values of the asymmetry parameter
$\theta$. For $L = 2, 3, 4, 5$ the results were obtained from exact
diagonalization.}
\end{figure}

\begin{figure}[th]
        \includegraphics[width=0.48\textwidth]{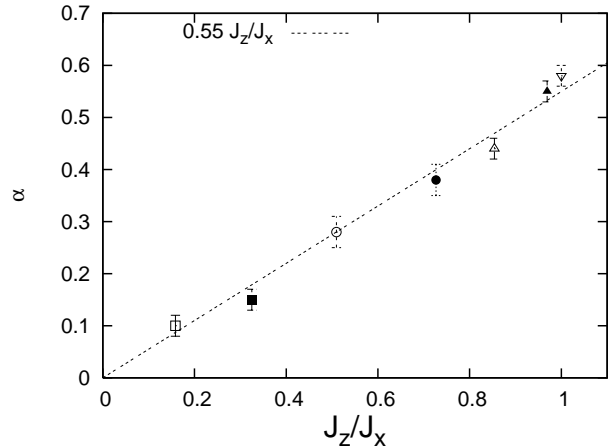}
        \caption{\label{fig_mc_plotScaling-theta}$\alpha$ versus
$J_z/J_x$, where $\alpha$ is the gap scaling constant $\Delta \sim
\alpha^L$. The symbols are the results obtained with Green's function
Monte Carlo and the dotted line is fitted to these values.}
\end{figure}

Fig.~\ref{fig_mc_plotGap-N_theta_excite} shows a log-linear plot of
the gap $\Delta$ versus the linear lattice size $L$ for various values of the
asymmetry parameter $\theta$.
Perturbation theory predicts that the scaling of $\Delta$ is given by
a power law $\Delta/J_x \propto \alpha^L$, with $\alpha \simeq 0.531
J_z / J_x$. To check this prediction, we have fitted the results of
Fig.~\ref{fig_mc_plotGap-N_theta_excite} with a straight line for each
value of $\theta$, keeping only sizes beyond which the scaling is
approximately linear. The values of $\alpha$ deduced from this fit are
plotted in Fig.~\ref{fig_mc_plotScaling-theta} as a function of
$J_z/J_x$.
Remarkably, the relation between $\alpha$ and $J_z / J_x$ is quite
linear up to $J_x = J_z$, which seems to justify the perturbation
theory in this limit. Moreover, a fit gives $\alpha \simeq 0.55
J_z/J_x$, in good agreement with the perturbation theory prediction
$\alpha \simeq 0.531 J_z/J_x$. In fact, for small $J_z/J_x$, even the
prefactor and the even-odd effect predicted by the perturbation theory
agree with the results of
Fig.~\ref{fig_mc_plotGap-N_theta_excite}. All these results lead us to
the conclusion that the gap $\Delta$ indeed follows a power law
$\Delta/J \sim \alpha^L$, with $\alpha < 1$ when $J_z \leq J_x$,
implying that $\Delta \rightarrow 0$ in the thermodynamic limit. Thus
the $2^L$ states coming from the limit $\theta = 0$ ($J_z = 0$) ground
state collapse in the thermodynamic limit, as long as $J_z\leq
J_x$. Using the rotation symmetry $R_y ( \frac{\pi}{2} )$, the same
result holds true for the $2^L$ states coming from the limit $\theta =
\frac{\pi}{2}$ (i.e., $J_x = 0$) ground state, as long as $J_z\geq
J_x$. The conclusion is that there are $2^L$ states collapsing
exponentially fast onto each other when $J_x\neq J_z$, and at least a
$2\times 2^L-2$ when (i.e., $J_x=J_z$). As we argued above, the actual
number is very probably actually equal to $2\times 2^L$, in agreement
with the semiclassical analysis.

\section{\label{sec_spin1}General spin}

The symmetry arguments used for spin-1/2 can be easily extended to
arbitrary spins.  Let us consider the system with $N$ spins $S$ on a
$L\times L$ lattice described by the compass model Hamiltonian
\begin{equation}
  H = - \sum_{\bm{r}} \left( J_x S_{\bm{r}}^x S_{\bm{r} + \bm{e}_x}^x 
      + J_z  S_{\bm{r}}^z S_{\bm{r} + \bm{e}_z}^z \right). \label{spin11}
\end{equation}

It is straightforward to check that the generalizations of the $P_j$
and $Q_i$ defined by
\begin{eqnarray}
  P_j & = & \prod_l i e^{- i \pi S_{l, j}^z} \\ 
  Q_l & = & \prod_j i e^{- i \pi S_{l, j}^x},
\end{eqnarray}
for integer spins and by
\begin{eqnarray}
  P_j & = & \prod_l  e^{- i \pi S_{l, j}^z} \\ 
  Q_l & = & \prod_j  e^{- i \pi S_{l, j}^x},
\end{eqnarray}
for half-integer spins commute with the Hamiltonian.  However, it is easy
to check that the commutator $[Q_i,P_j]$ vanishes for integer spins,
whereas it does {\it not} vanish for half-integer spins ($\{Q_i,P_j\}=0)$. 
Indeed, the only terms in the product which do not trivialy commute
are those on the site $(i,j)$. So we just have to show that $[e^{- i
\pi S_{i, j}^x},e^{- i \pi S_{i, j}^z}]=0$ for integer spins and
$\{e^{- i \pi S_{i, j}^x},e^{- i \pi S_{i, j}^z}\}=0$ for half-integer
spins. The operator $e^{- i \pi S_{i, j}^z}=R_z(\pi)$ corresponds to
the spin rotation by an angle $\pi$ about the $z$ axis acting at site
$(i,j)$. Applying this rotation to $e^{- i \pi S_{i, j}^x}$ gives 
\begin{eqnarray*}
R_z^{-1}(\pi)e^{- i \pi S_{i, j}^x}R_z(\pi)&=& \sum_n
\frac{(-i\pi)^n}{n!}R_z^{-1}(\pi)(S_{i, j}^x)^n R_z(\pi)\\
 &=& \sum_n
\frac{(-i\pi)^n}{n!}(-S_{i, j}^z)^n \\
 &=&e^{i \pi S_{i, j}^z}
\end{eqnarray*}
But for integer spins, $e^{i \pi S_{i, j}^z}=e^{-i \pi S_{i, j}^z}$,
while for half-integer spins, $e^{i \pi S_{i, j}^z}=-e^{-i \pi S_{i,
j}^z}$, which terminates the proof. This has two important
consequences.  First of all, in Dou\c{c}ot {\it et al.}'s argument,
the fact that $[Q_i,P_j]\ne 0$ was crucial to show that each state was
two-fold degenerate. We thus expect that for all integer spins, the
eigenstates are generically non-degenerate. Besides, for integer
spins, since all $Q_i$'s commute with all $P_j$'s, one can use all
these symmetries simultaneously, leading to $2^{2L}$ different
symmetry sectors.

\begin{figure}[th]
        \includegraphics[width=0.48\textwidth]{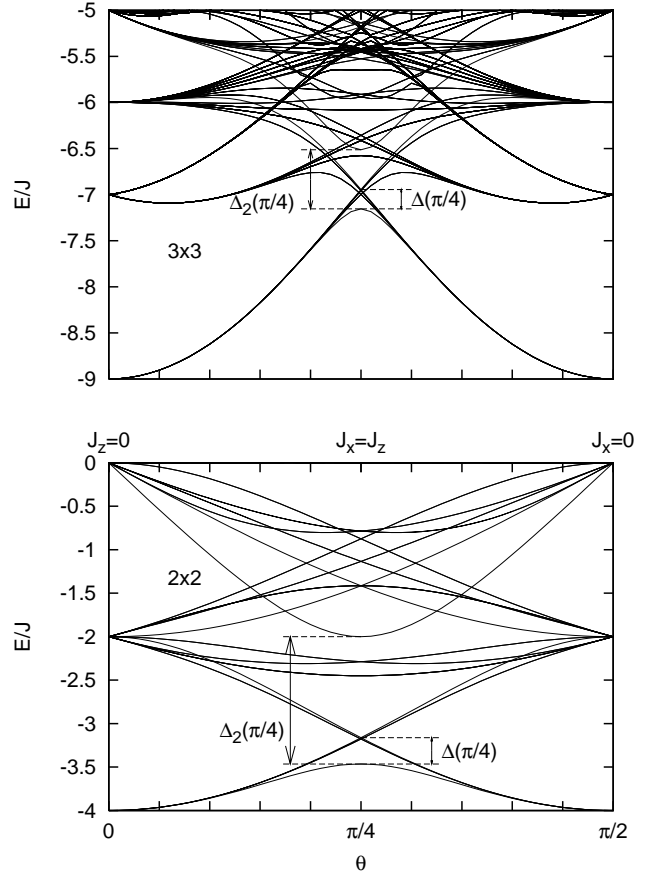}
        \caption{\label{fig_spin1_plot2x2_3x3_theta}Spectrum versus
asymmetry parameter $\theta$ for linear lattice sizes $L = 2$ and $L = 3$
obtained by exact diagonalization. $J_x = J \cos \theta$ and $J_z = J
\sin \theta$.}
\end{figure}

We have checked these predictions for $S=1$ with exact
diagonalizations of clusters of size $L=2$ and $L=3$ (see 
Fig.~\ref{fig_spin1_plot2x2_3x3_theta}).
Indeed, the ground state is non-degenerate for all $J_x,J_z\neq 0$,
and it is in the only symmetry sector $p_1 = \cdots = p_L = q_1 =
\cdots = q_L = ( - 1 )^L$ connecting the ground states in the limits
$J_x=0$ and $J_z=0$. The $J_z=0$ (resp.  $J_x=0$) ground state
$2^L$-fold degeneracy is partially lifted by the $J_z$ (resp. $J_x$)
term of the Hamiltonian, creating a gap $\Delta$. As in the spin-1/2
model, the gap between the ground state and the first excited state
which has the same degeneracy and is in the same symmetry sector as
the ground state is denoted by $\Delta_2$.  The gaps $\Delta$ and
$\Delta_2$ are smaller for $L=3$ than for $L=2$, and one can
conjecture that these gaps go to zero in the thermodynamic limit. In
this case, the ground state would have the same degeneracy as for
spin-1/2. A definite conclusion would clearly require to study larger
clusters though.

\section{Conclusion}

Using a variety of approaches, we have obtained a coherent picture of
the zero-temperature properties of the quantum compass model. On a
finite cluster, we have confirmed that all eigenstates of the spin-1/2
model are at least two-fold degenerate, a result that we have extended
to arbitrary half-integer spins, while they are not necessarily
degenerate for integer spins. However, the degeneracy that remains
when thermal or semi-classical quantum fluctuations are introduced,
namely the possibility to flip the spins along lines or columns, is
still present as a manifold of states which collapse exponentially
fast onto the ground state upon increasing the size of the lattice.
This was already known to be the case for the asymmetric case not too
close to $J_x=J_z$. Thanks to extensive quantum Monte Carlo
simulations, we have shown that this remains true up to the symmetric
case $J_x=J_z$, and that the the number of these states ($2^L$ when
$J_x\ne J_z$, $2\times2^L$ when $J_x=J_z$) agrees with the degeneracy
predicted by the semi-classical analysis.

Physically, this has two consequences. Regarding orbital fluctuations
in Mott insulators, in which case the symmetric version of the model
seems more appropriate, the present results confirm the absence of
true orbital long-range order for quantum spins in the thermodynamic
limit even at zero temperature. Nevertheless, as in the case of
thermal fluctuations for classical spins, the possibility to choose
between the $x$ and $z$ directions should still lead to a
finite-temperature Ising transition.

Regarding Josephson junction arrays, one of the important issues is to
ensure that the two-fold degenerate ground state is well protected by
a gap to all excited states. As noticed by Dou\c{c}ot {\it et al.},
this requires to work with not too large systems if a family of states
collapse onto the ground state in the thermodynamic limit, as they
already showed to be the case for $J_x\ll J_z$. Their results
suggested however that there might be a quantum phase transition to a
gapped phase around $J_x=J_z$ in which the gap to all excited states
would remain finite even in the thermodynamic limit. This possibility
is clearly ruled out by the present results.\\

\begin{acknowledgments}
We thank Mike Ma for enlightning discussions about the symmetries of
the model, Arnaud Ralko for useful discussions about Green's function
Monte Carlo, and Beno\^ \i t Dou\c{c}ot for a critical reading of the
manuscript.  This work was supported by the Swiss National Fund and by
MaNEP.  F.B. was supported by INFM and by MIUR (COFIN 2004).
\end{acknowledgments}


\begin{thebibliography}{1}
\bibitem{footnote} Unfortunately, this state is not accessible to the quantum 
Monte Carlo algorithm, which only gives access to the ground-state energy 
of a given sector.
\end{thebibliography}
\end{document}